\input harvmac
\sequentialequations
\def\Im{\rm Im}
\def\n#1{{\Im} N_{#1}}
\def\np#1{{\it Nucl. Phys.,} {\bf B#1},}
\def\prl#1{{\it Phys. Rev. Lett.,} {\bf #1},}
\def\prd#1{{\it Phys. Rev.,} {\bf D#1},}
\def\plb#1{{\it Phys. Lett.,} {\bf B#1},}
\def\tilde{\widetilde}

\nref\ferrara{S.~Ferrara, R.~Kallosh, and A.~Strominger 1995, ``N=2 
Extremal Black Holes," \prd{52} 5412-5416, hep-th/9508072.}
\nref\andy{A.~Strominger 1996, ``Macroscopic Entropy of N=2 Extremal Black 
Holes,'' \plb{383} 39-43, hep-th/9602111.}
\nref\kf{S.~Ferrara, and R.~Kallosh 1996, ``Supersymmetry and Attractors,'' 
\prd{54} 1514-1524, hep-th/9602136.}
\nref\renata{R.~Kallosh 1996, ``Superpotential from Black Holes,'' 
\prd{54} 4709-4713, hep-th/9606093.} 
\nref\ksw{R.~Kallosh, M.~Shmakova, and W.~K.~Wong 1996, 
``Freezing of Moduli by N=2 Dyons," \prd{54} 6284-6292, hep-th/9607077.}
\nref\gkk{G.~Gibbons, R.~Kallosh, and B.~Kol 1996, 
``Moduli, Scalar Charges, and 
the First Law of Black Hole Thermodynamics,'' \prl{77} 4992-4995, 
hep-th/9607108.}
\nref\bkrsw{K.~Behrndt, R.~Kallosh, J.~Rahmfeld, M.~Shmakova, and W.~K.~Wong 
1996,
``STU Black Holes and String Triality,'' \prd{54} 6293-6301, hep-th/9608059.}
\nref\clm{G.~L.~Cardoso, D.~Lust, and T.~Mohaupt 1996, ``Modular Symmetries 
of N=2 
Black Holes," \plb{388} 266-272, hep-th/9608099.}
\nref\behrndtC{ K.~Behrndt, G.~L.~Cardoso,
B.~deWit, R.~Kallosh, D.~Lust, and T.~Mohaupt 1996, ``Classical and 
Quantum N=2 Supersymmetric Black Holes," \np{488} 236-260, 
hep-th/9610105.}
\nref\sjrey{S.~J.~Rey 1996, ``Classical and Quantum Aspects of BPS Black 
Holes in 
N=2, D=4 Heterotic String Compactifications," hep-th/9610157.}
\nref\behrndtA{K.~Behrndt 1996, ``Quantum Corrections for D=4 Black Holes 
and D=5 Strings," \plb{396} 77-84,
hep-th/9610232.}
\nref\krw{R.~Kallosh, A.~Rajaraman, and W.~K.~Wong 1997, ``Supersymmetric 
Rotating Black Holes and Attractors,'' \prd{55} 3246-3249, hep-th/9611094.}
\nref\behrndtB{K.~Behrndt 1997, ``Classical and Quantum Aspects of 4 
Dimensional Black Holes," hep-th/9701053.}
\nref\shmakova{M.~Shmakova 1996, ``Calabi-Yau Black Holes,'' hep-th/9612076.}
\nref\bs{K.~Behrndt, and W.~A.~Sabra 1997, ``Static N=2 Black Holes for 
Quadratic 
Prepotentials,'' hep-th/9702010.}
\nref\sabra{W.~A.~Sabra 1997, ``General Static N=2 Black Holes," 
hep-th/9703101, ``Black Holes In N=2 Supergravity Theories And Harmonic 
Functions," hep-th/9704147.}

\lref\cvetic{M.~Cvetic, and A.~A.~Tseytlin 1996, ``Non-Extreme Black Holes 
from Non-Extreme Intersecting M-Branes," \np{478} 181-198, hep-th/9606033.}
\lref\tseytlin{A.~A.~Tseytlin 1996, ``Harmonic Superpositions of M-branes,'' 
\np{475} 149-163, hep-th/9604035.}
\lref\gkt{J.~Gauntlett, D.~Kastor, and J.~Traschen 1996, ``Overlapping Branes 
in 
M-Theory,'' \np{478} 544-560, hep-th/9604179.}
\lref\duff{M.~J.~Duff, H.~Lu, and C.~N.~Pope 1996, ``The Black Branes of 
M Theory," \plb{382} 73-80, hep-th/9604052.}
\lref\larsen{V.~Balasubramanian, F.~Larsen, and R.~G.~Leigh, ``Branes at 
Angles and Black Holes,'' hep-th/9704143.}
\lref\mirjam{M. Cvetic 1997, ``Properties of Black Holes in Toroidally 
Compactified String Theory,'' hep-th/9701152.}



\Title{\vbox{\baselineskip12pt
\hbox{UMHEP-442}
\hbox{hep-th/9705090}}}
{\vbox{\centerline{\titlerm Non-extreme Calabi-Yau Black Holes}
 }}

\centerline{ David Kastor
and K.~Z.~Win\footnote{}{kastor@phast.umass.edu,\ win@phast.umass.edu} }
\bigskip
\centerline{\it  Department of Physics and Astronomy}
\centerline{\it University of Massachusetts}
\centerline{\it Amherst, MA 01003-4525}

\bigskip

\bigskip
\centerline{\bf Abstract}
\bigskip
Non-extreme black hole solutions of four dimensional, $N=2$ 
supergravity theories with Calabi-Yau prepotentials 
are presented, which generalize certain known double-extreme and
extreme solutions.  
The boost parameters characterizing the nonextreme solutions must
satisfy certain constraints, which effectively limit the functional independence of
the moduli scalars.
A necessary condition for being able to take certain boost parameters 
independent is found to be
block diagonality of the gauge coupling matrix.  We present a number of examples aimed
at developing an understanding of this situation and speculate about the existence
of more general solutions.

\Date{May 1997}

\newsec{Introduction}  

\noindent
Considerable effort has been devoted recently to studying 
black hole solutions in four-dimensional, $N=2$ supergravity theories 
\refs{\ferrara-\sabra}. 
Interest has been focused, so far, on extreme black holes, which satisfy additional supersymmetry
constraints and saturate a BPS bound. A key discovery \kf\ in this case is 
that the values of the scalar moduli fields of the $N=2$ vector multiplets are 
actually fixed at the black
hole horizon in terms of the electric and magnetic charges carried by the black
hole.  In particular, the horizon values of the scalar fields are 
independent of the values of the scalar fields at infinity.  
The evolution of the scalar fields moving inward from
infinity towards the horizon can then be thought of as motion in a 
kind of attractor \kf.  Of particular interest are the ``double-extreme" 
solutions, for which the scalar fields stay fixed at their horizon values throughout the 
spacetime \behrndtC.  These are ``doubly" extreme in the sense that, 
in addition to having degenerate horizons, 
the black hole mass, for these solutions, is minimized for the given charges.  
``Singly" extreme solutions with non-constant scalars are given in 
\behrndtA .

In this paper we will look at non-extreme black hole solutions in $N=2$ theories in four 
dimensions, obtained by dimensional reduction of Type II supergravity on a Calabi-Yau threefold.
Since the basic form of the extreme solutions in this case \behrndtA\
is quite similar to certain supersymmetric, 
intersecting brane solutions of torus compactifications \refs{\tseytlin,\gkt}, 
a simple ansatz for the non-extreme $N=2$ black holes arises from
the known non-extreme intersecting brane solutions in torus compactifications 
\cvetic.
This ansatz is also analogous to the non-extreme generalization of the 
extreme
 black branes 
solution of M-theory \duff.
In this ansatz, given below, there is a single ``non-extremality" 
parameter $\mu$
 and 
a number of ``boost parameters"
$\gamma_\Lambda$ related to the individual charges.  
We find below, however, that this ansatz does not in
general solve the equations of motion.  Rather, the equations of motion 
reduce to a condition
which may be regarded as a constraint
on the boost parameters.  The only general 
({\it i.e.} for all Calabi-Yau manifolds) solution to this constraint, which we have found, 
is when all the boost parameters are taken to be equal.  For specific models, such 
as the $STU$ model and others discussed below, 
it is possible to take separate boost parameters.  

We have not yet explored these constraints fully. 
In the case of torus compactifications of $D=11$ 
supergravity, the general non-extreme solutions of \cvetic\ may be 
obtained from the $D=10$
Schwarzschild solution via various combinations of boosts, dimensional upliftings and 
reductions and duality symmetries.  We note that these same methods cannot be used to
similarly construct the non-extreme $N=2$ solutions.
\foot{After this work was completed, we 
found that the same ansatz for the non-extreme solutions had been made 
in \behrndtB. We
disagree with the claim there that the ansatz generally satisfies the equations of motion.}

\newsec{The Basic Setup: $N=2$ Lagrangian}

\noindent
We give only a brief summary of the formalism here.  A more complete 
treatment may be found in, {\it e.g.}, \behrndtC .  An $N=2$ supergravity 
theory in four 
dimensions includes, in addition to the graviton multiplet, $n_v$ vector 
multiplets and
$n_h$ hypermultiplets.  In our work we consistently take the hypermultiplet 
fields to be constant and will ignore them below.  The bosonic part
of the action is then given by\foot{We use the normalization 
$\epsilon_{\hat t\hat r\hat\vartheta\hat\varphi}=1$.}
\eqn\action{
S=\int d^4x\sqrt{-G}\left[ R-2g_{A\bar B}\partial_\nu z^A\partial^\nu 
\bar z^B-\hbox{${1\over 4}$} \left(F^\Lambda_{\mu\nu}F^{\Sigma\mu\nu} 
\n{\Lambda\Sigma} + 
F^\Lambda_{\mu\nu}{*F^{\Sigma\mu\nu}}{\rm 
Re}N_{\Lambda\Sigma}\right)\right], 
} 
where $G_{\mu\nu}$ is the spacetime metric, 
$z^A$ with $A=1,\dots ,n_v$ are complex scalar moduli fields parametrizing a 
special K\"ahler manifold and $F_{\mu\nu}^\Lambda =2\partial_{[\mu} 
A_{\nu]}^\Lambda$ with
 $\Lambda=0,1,\dots,n_v$ are the 
field strengths of $n_v+1$ $U(1)$ gauge fields $A^\Lambda_\mu$. 
Here, the complex scalars are related to the 
holomorphic symplectic sections $X^\Lambda$ by the inhomogeneous coordinates
 condition
\eqn\inhom{
z^A={X^A\over X^0}
}
The K\"ahler potential $K$, 
scalar metric $g_{A\bar B}$ and gauge couplings $N_{\Lambda\Sigma}$ are all
determined in terms of the
prepotential $F(X)$, which is a holomorphic, second-order homogeneous function.
The K\"ahler potential $K$ is given by 
\eqn\kahler{e^{-K}=i\left(\bar X^\Lambda F_\Lambda-X^\Lambda\bar 
F_\Lambda\right) } 
where $F_\Lambda= \partial F/\partial X^\Lambda$.
The K\"ahler metric on the scalar moduli space is then given by 
$g_{A\bar B}= \partial_A\partial_{\bar B}K(z,\bar z)$ where 
$\partial_{\bar A}=\partial/\partial{\bar z^A}$ and the gauge field couplings 
$N_{\Lambda\Sigma}$ by 
\eqn\gaugecoupling{
N_{\Lambda\Sigma}=\bar F_{\Lambda\Sigma}+2i({\Im} F_{\Lambda\Delta})({\Im} 
F_{\Sigma\Gamma})X^\Gamma X^\Delta/\left(X^\Omega X^\Phi{\Im} 
F_{\Omega\Phi}\right)
} 
where $F_{\Lambda\Sigma}=\partial F_\Lambda/\partial X^\Sigma$.

For type II supergravity compactified on a Calabi-Yau space, the prepotential takes the 
form
\eqn\prepotential{
F(X)={d_{ABC}X^AX^BX^C\over X^0},
}
where the constants $d_{ABC}$, with $ABC$ completely symmetric, 
 are the topological intersection numbers of the manifold.
We further restrict our interest here to the axion free case, in which  
all the moduli scalars $z^A$ are pure imaginary. 
The gauge coupling matrix $N_{\Lambda\Sigma}$ is then 
pure imaginary, having nonzero components 
\eqn\matrixN{
N_{00}=-d_{ABC}z^Az^Bz^C\quad , \quad N_{AB}=-6d_{ABC}z^C+
9{d_{ACD}z^Cz^Dd_{BEF}z^Ez^F\over d_{GHI}z^Gz^Hz^I}
}
and the K\"ahler 
metric is given by
\eqn\metric{
g_{A\bar B}={N_{AB}\over 4N_{00}}
}
The equations of motion following from the action  
(with ${\rm Re}N=0$) are given by
\eqnn\eomA
\eqnn\eomB
\eqnn\eomC
$$\eqalignno{
\partial_\mu\left(\sqrt{-G}F^{\Lambda\mu\nu}{\Im} 
N_{\Lambda\Sigma}\right)&=0 
&\eomA\cr 16g_{A\bar B}\nabla^\nu\partial_\nu\bar z^B
+8(\partial_A g_{B\bar C})\partial^\mu z^B\partial_\mu\bar z^C
-\left(\partial_A\n{\Lambda\Sigma}\right)F^\Lambda_{\mu\nu}F^{\Sigma\mu\nu}&=0&\eomB\cr
R_{\mu\nu}-2g_{A\bar B}(\partial_\mu z^A)\partial_\nu \bar 
z^B-\hbox{${1\over 2}$}\left(
F^\Lambda_{\mu\sigma}F^{\Sigma\sigma}_\nu-{g_{\mu\nu}\over 4}
F^\Lambda_{\rho\sigma}F^{\Sigma\rho\sigma}\right)\n{\Lambda\Sigma}&=0.&\eomC\cr
}
$$ 

\newsec{Non-Extreme Solutions}

\noindent 
We want to generalize certain double-extreme and extreme 
solutions, which were given in \behrndtC\ and  \behrndtA\ respectively.
  In these solutions, the 
gauge field $F_{\mu\nu}^0$ carries only electric charge, while each gauge 
field $F_{\mu\nu}^A$ 
  carries only magnetic charge.  As discussed in 
\refs{\behrndtC,\behrndtA}, regarded 
as a compactification of M-theory on $S^1\times CY$, these solutions correspond to 
fivebranes wrapping 4-cycles of the Calabi-Yau space, with a boost along the common string.
For the special case of a torus compactification, the corresponding 
non-extreme solutions are given in \cvetic .
It is straightforward to modify the solutions there to get an  
ansatz for the non-extreme solutions in the present case,
\eqnn\nonextreme
$$\eqalign{ds^2=-e^{-2U}fdt^2+e^{2U}\left(f^{-1}dr^2+r^2d\Omega^2\right)
\quad,&\quad 
e^{2U}=\sqrt{H_0d_{ABC}H^AH^BH^C}\cr
 f=1-{\mu\over r}\quad,\quad
z^A=iH^AH_0e^{-2U}\quad,&\quad H^A=h^A\left(1+{\mu\over 
r}\sinh^2\gamma_A\right)\cr
A^0_t={r\tilde H_0'\over h_0 H_0}\quad,\quad 
A^C_\varphi=r^2\cos\vartheta\ 
\tilde H^{C^\prime}\quad,&\quad\tilde H^A=h^A\left(1+{\mu\over 
r}\cosh\gamma_A\sinh\gamma_A\right)\cr
H_0=h_0\left(1+{\mu\over r}\sinh^2\gamma_0\right)\quad ,&\quad
\tilde H_0=h_0\left(1+{\mu
\over r}\cosh\gamma_0\sinh\gamma_0\right),\cr
}\eqno\nonextreme$$
where prime denotes $\partial_r$. Nonzero components of the gauge field strengths are
\eqn\strength{
F^0_{tr}={\tilde H_0'\over H_0^2}\quad,\quad F^A_{\varphi\vartheta}=
r^2\sin
\vartheta
\tilde H^{A^\prime}.}
The ansatz \nonextreme\ reduces to the ``singly'' extreme 
solutions
given in
\behrndtA\ when the limit $\mu\rightarrow 0, 
\gamma_\Lambda\rightarrow\infty$ is taken
 with
 $\mu\sinh^2\gamma_\Lambda\equiv k_\Lambda$ held fixed and further to the ``doubly'' 
extreme solutions, with constant moduli scalars, in \behrndtC\ when all the 
$k_\Lambda$ are the same.
It can also be shown that, if the solution \nonextreme\ 
with $H_0=\tilde H_0=1$ satisfies the equations of motion, then the solution  
with more general $H_0$ and $\tilde H_0$, as given in \nonextreme,
satisfies the equations of motion. 
This corresponds to a boost transformation
 in M-theory 
compactified on $S^1\times CY$.
Henceforth, in checking the equations of motion, we will set $H_0=\tilde H_0=1$.
  
It is straightforward to check that the ansatz \nonextreme\ satisfies the
gauge field equation of motion \eomA .  Equation \eomC\ for the curvature 
reduces to the condition
\eqn\eqA{
r^2\n{AB}
\left( f{H^A}'{H^B}'-
{\tilde H}^{A^\prime}
{\tilde H}^{B^\prime}
\right)=2\mu\left(e^{2U}\right)', 
} 
and the scalar field equation \eomB\ leads to 
\eqn\eqB
{
r^2\left(\partial_A\n{BC}\right)
\left(
{\tilde H}^{C^\prime}{\tilde H}^{B^\prime}
-f{H^C}'{H^B}'\right)=8\mu e^{2U}g_{A\bar B}{\bar z}^{B^\prime}.
}
In deriving these last two equations we have made use of the fact that the 
extreme
solutions, with $f=1$ and $(H_0,H^A)=(\tilde H_0,\tilde H^A)$, satisfy the 
equations of motion.
Note that both sides of equations \eqA\ and \eqB\ vanish identically in this case.  
We also note that $\n{BC}=-iN_{BC}$ by virtue of \matrixN\ 
is a first order homogeneous function of $z^A$ and that, in  
particular, $z^A\partial_A\n{BC}=\n{BC}$.  This property can be used to  
``contract" equation \eqB\ with $z^A$ to obtain equation \eqA.  Thus it is only 
necessary to show that the ansatz
\nonextreme\ (with $H_0=\tilde H_0=1$) satisfies \eqB.

It is not difficult to see that,
for an arbitrary choice of the constants $d_{ABC}$ in the 
prepotential, the condition \eqB\ is not satisfied unless the parameters $\gamma_A$ are taken to
be equal. This differs from the case of intersecting branes on a torus \cvetic, 
for parameters $\gamma_A$ may be specified independently for each set of branes.  
We do not 
at present fully understand the significance of the restrictions placed by \eqB\ on the 
parameters
$\gamma_A$.  Note that, if all the boost parameters, including $\gamma_0$, are set equal 
to some common value $\gamma$ in \nonextreme, then the scalars $z^A$ 
will be constant, having values
\eqn\constants{z^A=ih^Ah_0,}
where the asymptotic flatness condition, $h_0d_{ABC}h^Ah^Bh^C=1$, 
has been used.  
This case is then a non-extreme version of the ``doubly'' extreme black holes in \behrndtC.
Taking $\gamma_0$ to be different, as may always be done, makes the scalars $z^A$ non-constant, but keeps their
 ratios constants.  Clearly, if some, or all, of the $\gamma_A$'s may 
also be taken unequal, 
then there will be additional functional independence between the scalars.
In the next section, we will explore some simple examples of prepotentials for which 
some, or all, of the $\gamma_A$'s may be specified independently.

\newsec{Examples}

\noindent 
We list below  some choices for the $d_{ABC}$ which  allow some of the
$\gamma_A$'s also to be different from each other.  
It follows from \eqA , that a
{\it necessary} condition for (at least) some of the $\gamma_A$'s to be independent is that the
gauge coupling matrix $\n{AB}$ be block diagonal.  
In this case there turns out to be
one independent parameter per block.  From this point of view, it seems consistent that
$\gamma_0$ may always be specified independently of $\gamma_A$, since $N_{0A}$ vanishes as evident
by \matrixN, and hence $N_{00}$ forms a $1\times 1$ block.

Our first example is the $STU$ model $\behrndtC$ for which the only 
nonzero $d_{ABC}$ is $d_{123}$.
In this case the coupling matrix $\n{BC}$ is diagonal and all three 
parameters
$\gamma_1,\gamma_2,\gamma_3$ may all be specified independently.  
However, when quantum corrections are added to the $STU$ model 
\refs{\behrndtC,\behrndtA}\ 
$d_{333}$ becomes nonzero.  This makes the coupling matrix  
$\n{BC}$ completely nondiagonal, which in turn implies that the $\gamma_A$'s 
must be taken equal.

As a second example, we can take only the constants $d_{1AB}$ to be nonzero, where 
$A,B\neq 1$ (a similar model is considered in \ksw).  The coupling matrix $\n{BC}$ in this case 
is block diagonal, having a $1\times 1$ block and an $(n_v-1)\times (n_v-1)$ 
block.  It follows that $\gamma_1$ can be chosen independently of the 
$\gamma_A$ for $A\ne 1$, which must all be the same. 

A specialization of the previous example is to take 
only $d_{12B}$ nonzero with 
$B=3\dots n_v$. This  makes $\n{BC}$ block diagonal 
with two  $1\times 1$ blocks and one $(n_v-2)\times (n_v-2)$ block and 
one can have three different $\gamma$'s: $\gamma_1$, $\gamma_2$ and 
one more $\gamma_B$ for $B=3\dots n_v$. 
\foot{Notice that if one specializes this last example one step further one 
ends up with the $STU$ model (without the quantum correction).}

As a final example we consider a simple toy model
 where only $d_{112}$ and 
$d_{111}$ are nonzero.
In this case $\n{BC}$ is diagonal if and only if 
$d_{111}=0$, {\it i.e.} $\gamma_1=\gamma_2$ is required unless 
$d_{111}=0$.     
In each of these cases block diagonality of the gauge coupling matrix 
$\n{BC}$ appears to be both a necessary and a
sufficient condition to be able to take independent $\gamma$'s, though we have not been able
to show this generally.

\noindent

\newsec{Physical Parameters and Discussion}

\noindent
We examine the physical properties of the non-extreme 
solutions.
In particular, we want to check, given the restrictions on the $\gamma_A$'s,
that the charges may still be specified arbitrarily, as they can 
in the extreme limit \refs{\behrndtC,\behrndtA}.
We will first display all formulae as if the $\gamma_A$'s can be specified 
independently and then discuss the actual solutions, in which the $\gamma_A$'s are restricted.
After imposing the asymptotic flatness 
condition, the set of independent parameters for the solutions can be taken to be
$\{\mu,\gamma_0,h^A,\gamma_A\}$. These can be exchanged for the more physical set 
$\{E,q_0,p^A,\gamma_A\}$, where $E$ is the ADM mass, $q_0$ the electric charge 
for $F_{\mu\nu}^0$ and $p^A$ the
magnetic charges for $F_{\mu\nu}^A$.
The ADM energy is given by\foot{In order to simplify the formulae we  explicitly
display $h_0$ bearing in mind that it can be regarded as a function of $h^A$.}
\eqn\energy{
E=\hbox{${1\over 2}$}\left[\mu+\hbox{${1\over 2}$}
\left(
k_0+
3h_0d_{ABC}h^Ah^B{\cal K}^C
\right)\right]
}
where ${\cal K}^C\equiv h^Ck_C$ and $k_\Lambda=\mu\sinh^2\gamma_\Lambda$ as above.
The electric charge $q_0$ and magnetic charges $p^A$ are defined by 
\eqn\chargeintegrals{
q_0={1\over 4\pi}\int{*F^0_{\vartheta\varphi}}\n{00}\ 
d\vartheta d\varphi \ ,\quad
p^A={1\over 4\pi}\int F^A_{\vartheta\varphi} \ d\vartheta d\varphi .}
We find 
\eqn\charges{
q_0={\mu h_0\sinh 2\gamma_0\over 2}
\ ,\quad 
p^A={\mu h^A\sinh 2\gamma_A\over 2}}
The Hawking temperature is  
\eqn\temperature{
 T={1\over
4\pi\mu\sqrt{
\lambda_0d_{ABC}\lambda^A\lambda^B\lambda^C}}
}
where $\lambda_0=h_0\cosh^2\gamma_0$ and $\lambda^A=h^A\cosh^2\gamma_A$ and 
the Bekenstein entropy is 
\eqn\entropy{S=\pi\mu^2\sqrt{
\lambda_0d_{ABC}\lambda^A\lambda^B\lambda^C}.}

First, note that equation \charges\ implies that, 
even in the case that all boost parameters are set equal, 
the charges $q_0,p^A$ may still be chosen arbitrarily by virtue of the constants $h^A$ and
the single boost parameter $\gamma$.  
As we observed above, the restrictions on the 
$\gamma_A$ should be regarded as restrictions on the functional independence of the 
scalars $z^A$, with respect to one another.
Next, we note that, for all the examples  discussed in the last section, the 
formulae for
the temperature \temperature\ and the entropy \entropy\ simplify considerably.
The  square roots in \temperature\ and \entropy\ can be ``gotten rid of", in these cases,
because the $\lambda$ factors
appearing in the each term of the sums are identical.
For example,  in the
$d_{1AB}$ model, the entropy \entropy\ reduces to
\eqn\modelentropy{
S=\pi 
\mu^2\cosh\gamma_0\cosh\gamma_1\cosh^2\gamma ,}
 where $\gamma=\gamma_A$ for 
$A=2\dots n_v$.

It remains an open question, whether, or not, more general non-extreme solutions (static, 
axion-free and carrying only the charges $q_0$ and $p^A$) exist.  
These might, for example, have independent boost parameters for each of the Calabi-Yau 
$4$-cycles.  In the case of orthogonally intersecting branes on a torus \cvetic, there are
at most four independent parameters corresponding to a boost and three sets of branes.  However,
the most general black hole solutions in type II theory compactified to $4$-dimensions on a 
torus are described by $28$ electric and $28$ magnetic charges (see {\it e.g.} \mirjam ).  
The extreme solutions in this case arise via collections of branes intersecting 
{\it non-orthogonally} \larsen.  
It may be necessary to look at a non-extreme solution based on branes intersecting at
angles to get the most general solution in the Calabi-Yau case as well.
It would also be interesting to try to construct the solutions, which we have found here, 
using the available symmetry
transformations, which in the present case include boosts in the time direction and 
symplectic transformations.

Finally, it should also be possible to find nonextreme solutions in $N=2$ theories with
prepotentials not of the Calabi-Yau form.
We note that since \eqA\ and \eqB\ are derived using 
the extreme solution and since they 
are displayed not in 
terms of the particular prepotential we have used in this paper, they are 
generally applicable to finding non-extreme black hole solutions
for other 
prepotentials.  In particular  the block diagonality of $\n{AB}$ is
a  {\it necessary} condition for the existence of more than one $\gamma_A$.
We emphasize that the derivation of \eqA\ and \eqB\ does not depend on any 
specific expression for $e^{2U}$ and depends only on 
the fact that ${\rm Re}N=0$,  $F^0_{\mu\nu}=0$, and $F^A_{\mu\nu}$ carries 
only magnetic charge.

\listrefs
\bye